\documentclass[aps,prl,twocolumn,superscriptaddress,showpacs]{revtex4}

\usepackage{graphicx}
\begin{document}


\title{Anomalous metallic state in the vicinity of Metal to Valence Bond Solid insulator transition in LiVS$_2$}

\author{N. Katayama} \email{nk2u@cms.mail.virginia.edu}
\thanks{Present address: Department of Physics, University of Virginia, 382 McCormick Road, Charlottesville, VA 22904, USA}
\affiliation{Department of Advanced Materials, University of Tokyo, Kashiwa 277-8561, Japan}
\author{M. Uchida}
\affiliation{RIKEN (The Institute of Physical and Chemical Research), Wako 351-0198, Japan}
\author{D. Hashizume}
\affiliation{RIKEN (The Institute of Physical and Chemical Research), Wako 351-0198, Japan}
\author{S. Niitaka}					
\affiliation{RIKEN (The Institute of Physical and Chemical Research), Wako 351-0198, Japan}
\author{J. Matsuno}
\affiliation{RIKEN (The Institute of Physical and Chemical Research), Wako 351-0198, Japan}
\author{D. Matsumura}
\affiliation{Synchrotron Radiation Research Center, Japan Atomic Energy Agency, Hyogo 679-5148, Japan}
\author{Y. Nishihata}
\affiliation{Synchrotron Radiation Research Center, Japan Atomic Energy Agency, Hyogo 679-5148, Japan}
\author{J. Mizuki}
\affiliation{Synchrotron Radiation Research Center, Japan Atomic Energy Agency, Hyogo 679-5148, Japan}
\author{N. Takeshita}
\affiliation{Correlated Electron Research Center (CERC), National Institute of Advanced Industrial Science and Technology (AIST), Tsukuba 305-8562, Japan}
\author{A. Gauzzi}
\affiliation{Department of Advanced Materials, University of Tokyo, Kashiwa 277-8561, Japan}
\affiliation{Institut de Min$\acute{e}$ralogie et de Physique des Milieux Condens$\acute{e}$s-CNRS, Universit$\acute{e}$ Pierre et Marie Curie-Paris 6, 4, place Jussieu, 75252, Paris, France}
\author{M. Nohara} 
\thanks{Present address: Department of Physics, Okayama University, 3-1-1 Tsushima-naka, Okayama 700-8530, Japan}
\affiliation{Department of Advanced Materials, University of Tokyo, Kashiwa 277-8561, Japan}
\author{H. Takagi}
\affiliation{Department of Advanced Materials, University of Tokyo, Kashiwa 277-8561, Japan}
\affiliation{RIKEN (The Institute of Physical and Chemical Research), Wako 351-0198, Japan}
\affiliation{Correlated Electron Research Center (CERC), National Institute of Advanced Industrial Science and Technology (AIST), Tsukuba 305-8562, Japan}

\date{\today}

\begin{abstract}
We investigate LiVS$_2$ and LiVSe$_2$ with a triangular lattice as itinerant analogues of LiVO$_2$, known for the formation of valence bond solid (VBS) state out of $S$ = 1 frustrated magnet. LiVS$_2$, which is located at the border between a metal and a correlated insulator, shows a first ordered transition from a paramagnetic metal to a VBS insulator at $T_c$ $\sim$ 305 K upon cooling. The presence of VBS state in the close vicinity of insulator-metal transition may suggest the importance of itinerancy in the formation of VBS state. We argue that the high temperature metallic phase of LiVS$_2$ has a pseudo-gap, likely originating from the VBS fluctuation. LiVSe$_2$ was found to be a paramagnetic metal down to 2 K.
\end{abstract}

\maketitle

Transition metal compounds with a geometrically frustrated lattice, such as a triangular and a pyrochlore lattice, often form a valence bond solid (VBS) state at low temperatures. When the $t_{2g}$ orbitals are partially occupied, utilizing orbital degrees of freedom, complex ``molecular"  clusters in a spin singlet state are often formed: trimer in LiVO$_2$ \cite{rf:Goodenough,rf:LiVO2-NMR,rf:Onoda,rf:Imai,rf:LiVO2-1}, heptamer in AlV$_2$O$_4$ \cite{rf:AlV2O4-1}, helical dimer in MgTi$_2$O$_4$ \cite{rf:MgTi2O4} and octamer in CuIr$_2$S$_4$ \cite{rf:CuIr2S4-1}. 
A VBS state was found also in organic systems with geometrical frustration \cite{rf:Tamura}. 
Very recently, Shimizu {\it et al.} demonstrated that the VBS state can be melted by applying hydrostatic pressure $P$ on the organic compound
EtMe$_3$P[Pd(dmit)$_2$]$_2$ with a triangular lattice. Remarkably, superconductivity appears as soon as the VBS state is suppressed by $P$ \cite{rf:Shimizu}. 

A question that arises is whether or not similar melting of the VBS state 
and appearance of exotic metallic phases can occur in inorganic frustrated systems. In the inorganic systems, however, application of an external pressure is expected not to melt but to stabilize VBS due to a predominant volume effect. In CuIr$_2$S$_4$, the lattice shrinks appreciably in the VBS perhaps due to the formation of strongly bonded singlet molecules and the VBS can be stabilized through -$pV$ ($p$ : pressure, $V$ : volume) term in the corresponding free energy \cite{rf:Kobayashi}\cite{rf:CuIr2S4-P}. Effects of {\it negative} pressure on the VBS states of inorganic systems, on the other hand, have not been investigated so far.

The inorganic LiVO$_2$ in which the magnetic V$^{3+}$ ions ($3d$$^{2}$, $S$ = 1)  form a triangular lattice is known to be a paramagnetic insulator with strong antiferromagnetic interactions between the localized $S$ = 1 moments at high temperatures. Upon cooling, at $T_c$ $\sim$ 500 K, LiVO$_2$ exhibits a first ordered phase transition to a VBS state with a characteristic spin gap of $\sim$ 1600 K, evidenced by the formation of vanadium trimers. With this system, one can apply ``negative" pressure by replacing oxygens with larger anions such as S and Se \cite{rf:LiVS2-1,rf:LiVS2-2,rf:structure}. More over, the negative pressure may increase the overlap between V $3d$ and $p$-orbital (O $2p$, S $3p$, and Se $4p$), and increase the electronic band width. Thus, this vanadium- based triangular system provides a good opportunity to study effects of negative pressure on VBS states in inorganic materials.
In this Letter, we demonstrate that LiVS$_2$ is indeed an itinerant analogue of LiVO$_2$ with suppressed VBS. We found
that in LiVS$_2$ a phase transition from a paramagnetic metal to
a trimer VBS insulator occurs at $T_c$ $\sim$ 305 K that is lower
than that of LiVO$_2$. In LiVSe$_2$ with highest negative pressure,
the phase transition is suppressed down to 2 K. In the high
temperature metallic phase of LiVS$_2$, strong temperature dependence
of the bulk susceptibility, $\chi$, was observed, which is similar
to the pseudo-gap behavior found in underdoped superconducting
cuprates. We argue this is an evidence for a pseudo-gap formation by
short-range spin singlet fluctuations in the paramagnetic metallic
phase of LiVS$_2$.

Powder samples of LiVS$_2$, LiVSe$_2$ and their solid solution LiVS$_{2-x}$Se$_x$  
were prepared by a soft-chemical method followed by a solid-state reaction. 
For LiVS$_{2-x}$Se$_x$ (0 $\leq$ $x$ $\leq$ 0.4), 
Li-deficient Li$_{\sim0.75}$VS$_{2-x}$Se$_x$ was obtained first, 
by a reaction of an appropriate amount of Li$_2$S, V, S and Se 
in an evacuated quartz tube at 700 $^\circ$C for 3 days. 
For LiVSe$_2$, VSe$_2$ was synthesized from an appropriate amount of 
V and Se at the same condition with Li$_{\sim0.75}$VS$_{2-x}$Se$_x$. 
The products were immersed in 0.2 M $n$-BuLi hexane solution for 4 days to attain the maximum Li content \cite{rf:LiVS2-1}.
%
The samples were characterized by powder x-ray diffraction. 
The electron diffraction measurements were carried out in a HF-3000S (Hitachi) transmission electron microscope. 
Differential scanning calorimetry (DSC) was conducted by using DSC 204 F1 Phoenix (Netzsch).
Vanadium K-edge extended x-ray absorption fine structure (EXAFS) was measured at BL14B1, SPring-8.
Magnetic susceptibility was measured by a SQUID magnetometer (Quantum Design). 
Electrical resistivity was measured by a four-probe method. 
The powder samples were sintered at 500 $^\circ$C under Ar atmosphere for the resistivity measurements. 

\begin{figure}
\center
\includegraphics[width=8.8cm]{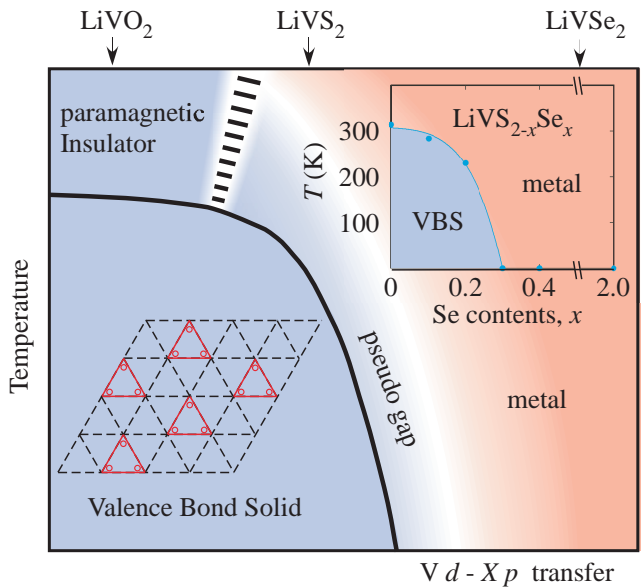}
\caption{\label{fig:phase}
(Color online) Schematic phase diagram in LiVO$_2$, LiVS$_2$ and LiVSe$_2$ system. Spin pseudo-gap is observed in the white region in the metallic phase. 
The left inset shows schematic VBS state on the triangular lattice of V$^{3+}$. The red circles denote the V ions.
The right inset shows the phase diagram in the vicinity of the VBS transition. Filled circles denote the VBS transition obtained from magnetic measurements for the solid solution LiVS$_{2-x}$Se$_x$.}
\end{figure}

\begin{figure}
\center
\includegraphics[width=8.8cm]{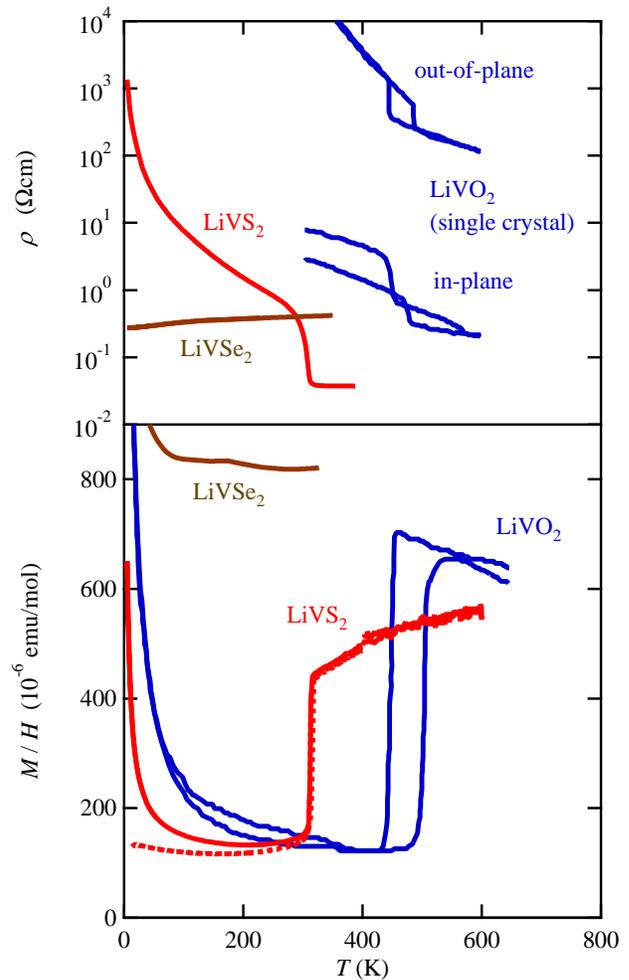}
\caption{\label{fig:graph}
(Color online) (a) Electrical resistivity and (b) magnetic susceptibility of LiVO$_2$, LiVS$_2$ and LiVSe$_2$ as a function of temperature. The electrical resistivity and magnetic susceptibility data of LiVO$_2$ are cited from ref.[5]. 
The broken line in magnetic susceptibility has been corrected for paramagnetic impurities as explained in the text.
}
\end{figure}

LiVS$_2$ exhibits a first ordered metal to insulator transition at $T_c$ $\sim$ 305 K, shown in Fig.\ref{fig:graph}. At high temperatures above $T_c$, the resistivity is about 40 m$\Omega$cm and almost temperature independent. Since the sample is low temperature sintered polycrystal, empirically, the intrinsic resistivity can be more than 1 order of magnitude smaller than 40 m$\Omega$cm, consistent with the metallic nature. Accompanied with the metal to insulator transition, an abrupt decrease in the magnetic susceptibility is observed, as shown in Fig.\ref{fig:graph}. The system is very likely to be nonmagnetic below $T_c$ with a temperature-independent Van Vleck term and a tiny low-temperature Curie tail, which corresponds to paramagnetic impurities of $\sim$ 1 \% if we assume spin 1/2 moment. In accord with the nonmagnetic behavior of LiVS$_2$, $^{51}$V NMR relaxation rate $T_{1}^{-1}$ shows thermally activated behavior, from which we estimate a spin gap of $\Delta$ = 1900 K \cite{rf:LiVS2-NMR}.



\begin{figure}
\center
\includegraphics[width=8.8cm]{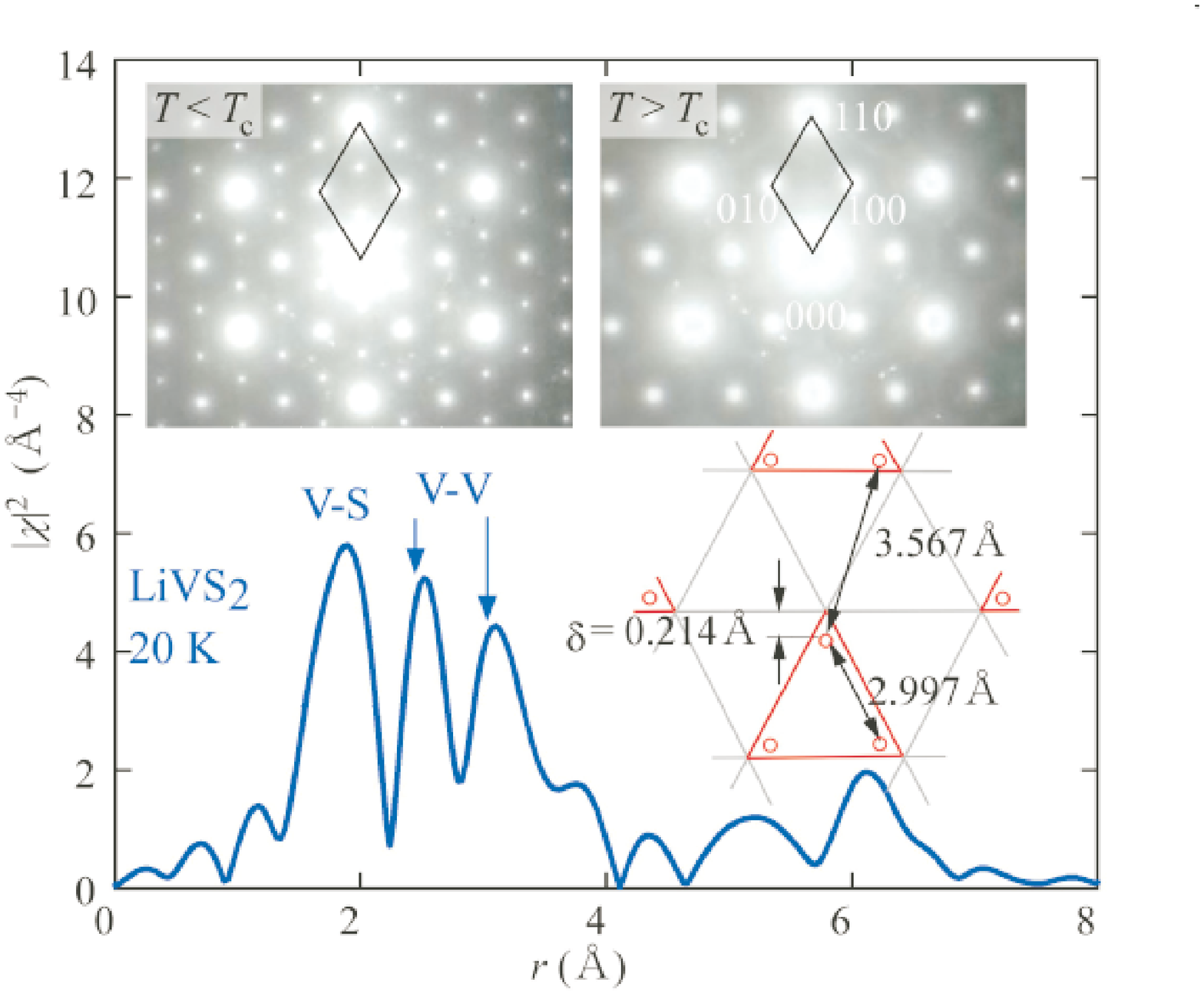}
\caption{\label{fig:EXAFS}
(Color online) Fourier transforms of the EXAFS spectra of LiVS$_2$ at 20 K. The lower right inset shows vanadium trimer schematically. The V-V distances and the displacement of the V from the regular triangular lattice, $\delta$, are determined from the data taken at 200 K.
The upper panels show the electron diffraction pattern along the [001] zone axis of LiVS$_2$.}
\end{figure}

Despite the metallic behaviors above $T_c$, electron diffraction measurements
on LiVS$_2$ show an evidence for the formation of the V trimers below
$T_c$ which indicates development of the same VBS state as in the
insulating LiVO$_2$.
The electron diffraction pattern reveals sharp superlattice reflections at $\{1/3~1/3~0\}$ below $T_c$ $\sim$ 305 K, as in Fig.\ref{fig:EXAFS}.  The superlattice reflections correspond to a $\sqrt{3}a\times\sqrt{3}a$ superlattice in real space, suggesting a formation of vanadium trimers in the VS$_2$ plane, shown in the right inset of Fig.\ref{fig:EXAFS}, The Fourier-transformed patterns of EXAFS spectra, shown in Fig.\ref{fig:EXAFS}, is indeed consistent with the vanadium trimers in low temperature phase. Below $T_c$, spectra shows clear three peaks between 1.5 and 3.5 \AA. The first peak at around 2 \AA\ is ascribed to that from the first-neighbored V-S. The second and third peaks, marked by arrows, are ascribed to those from the first-neighbored V-V bonds, indicating the presence of two inequivalent V bonds as expected for V trimer formation. Note the large splitting of more than 10 \%, indicative of a local character of V trimer.

\begin{figure}
\center
\includegraphics[width=8cm]{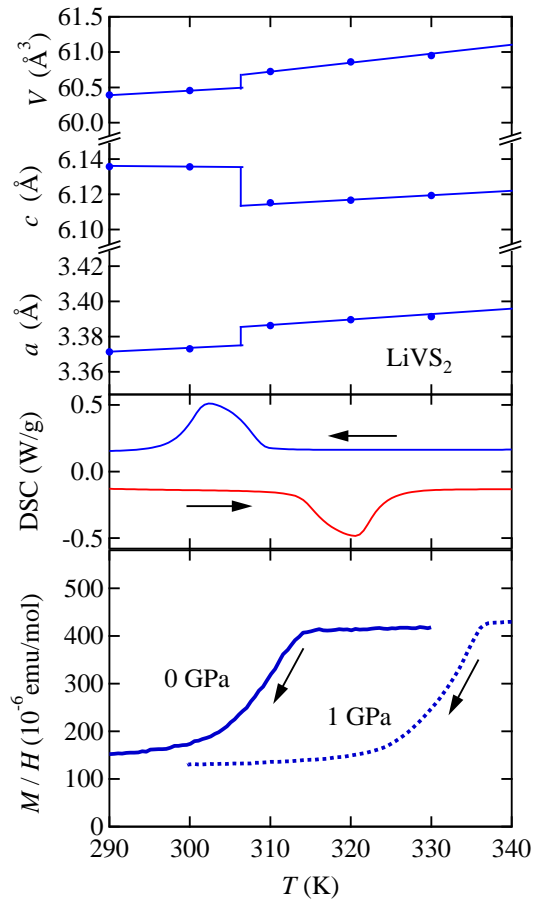}
\caption{\label{fig:rho_chi-2}
(Color online) The upper panel shows temperature dependence of lattice parameters $a$, $c$ and unit cell volume $v$. The middle panel shows differential scanning calorimetry (DSC) signals of LiVS$_2$. The lower panel shows magnetic susceptibility data of LiVS$_2$ under hydrostatic pressure. The measurements of lattice parameters and magnetic susceptibility were performed on cooling process, while DSC measurements were performed on both cooling and heating processes as indicated by arrows.}
\end{figure}

Associated with the formation of VBS state with V trimers,
the volume contraction originates from the large in-plane contraction.
We observed the volume contraction of $\sim$ 0.3 \% as shown in Fig.\ref{fig:rho_chi-2}, which is a factor of two smaller than that of LiVO$_2$ \cite{rf:Kobayashi}.
The DSC measurement shown in Fig.\ref{fig:rho_chi-2} indicates the increase of entropy $\Delta$$S$ $\sim$ 6.4 J/mol K
at the VBS transition. This should give rise to a positive pressure coefficient of VBS transition temperature
$dT_c$/$dP$ = $\Delta$$V$/$\Delta$$S$ $\sim$ 20 K/GPa from Clausius-Clapeyron relationship.
Indeed, we observe an increase of $T_c$ under an external pressure of a comparable magnitude in Fig.\ref{fig:rho_chi-2},
which implies the pressure induced stabilization of VBS simply reflects the volume contraction and
the low entropy in the VBS phase.


The high temperature metallic phase above $T_c$ $\sim$ 305 K in LiVS$_2$,
namely realized at the close vicinity to VBS, is not a simple metal but a pseudo-gap metal.
The magnetic susceptibility in this high temperature metallic phase is $\sim$ 5 $\times$ 10$^{-4}$ emu/mol,
which yields an estimate of electronic specific heat gamma $\gamma$ $\sim$ 35 mJ/mol K$^2$,
assuming the Wilson Ration $R_W$ $\sim$ 1. Since $R_W$ in correlated metals are normally 1.5-2.0,
a better estimate for gamma may be $\gamma$ $\sim$ 20 mJ/mol K$^2$. This is indeed consistent
with the entropy change $\Delta$$S$ $\sim$ 6.4 J/mol K at the VBS transition.
This $\Delta$$S$ for LiVS$_2$ is smaller than that for its insulating analogue LiVO$_2$
($\Delta$$S$ $\sim$ 14.5 J/mol K \cite{rf:Niitaka}), indicating that high temperature metallic state of LiVS$_2$
has lower entropy than the high temperature phase of LiVO$_2$,
which may be reasonably ascribed to a Fermi degeneracy in the metallic phase. Considering that $\Delta$$S$ for LiVO$_2$ is only roughly twice of $\Delta$$S$ for LiVS$_2$,
LiVS$_2$ may locate close to LiVO$_2$ in the phase diagram, shown in Fig.\ref{fig:phase}.
Assuming the observed $\Delta$$S$ in LiVS$_2$ originates from the electronic gamma $\gamma$ $\sim$ 21 mJ/mol K$^2$,
which is in good agreement with the estimate from the magnetic susceptibility.
These point that LiVS$_2$ at high temperatures is a paramagnetic metal with a moderate
density of states at the Fermi level, comparable to those of other 3$d$ transition metal chalcogenides.

In sharp contrast to the moderate magnitude, the temperature dependence of magnetic susceptibility is anomalous as a paramagnetic metal. 
As clearly seen in Fig.\ref{fig:graph}, the magnetic susceptibility shows a rapid decrease on cooling temperature, 
which is reminiscent of those of under-doped high-$T_c$ cuprates with pseudo-gap.
The same behavior of susceptibility was observed for powder sample previously \cite{rf:LiVS2-high}.
By replacing S with Se in LiVS$_2$, we can suppress the VBS and increase the band width further.
As shown in Fig.\ref{fig:graph}, we indeed find that LiVSe$_2$ is a paramgnetic metal down to
2 K without any trace of anomalous decrease of susceptibility, implying that the decrease of susceptibility in LiVS$_2$ is
associated with the proximity to the VBS state and metal-insulator transition.
The systematic evolution of electronic states on going from O, S to Se is clear and can be summarized
by a schematic phase shown in Fig.\ref{fig:phase}.
Note the increased magnitude of susceptibility in LiVSe$_2$ compared with LiVS$_2$ despite the increased band width.
This implies that the density of states of LiVS$_2$ is suppressed by some mechanism, which we take evidence of pseudo-gap.
In the electron diffraction pattern of LiVS$_2$ in the high temperature phase (Fig.\ref{fig:EXAFS}),
we observe a diffuse scattering indicative of short-range (and dynamic) trimer formation.
This leads us to speculate that the origin of pseudo-gap in the metallic LiVS$_2$ is a singlet fluctuation.

As summarized in Fig.\ref{fig:phase}, VBS state robustly appears from the insulating side to the metallic side.
VBS eventually vanishes for LiVSe$_2$ perhaps due to a combined effect of the increased band width and the lattice expansion. 
VBS formation in LiVO$_2$ had been discussed in terms of novel interplay of spin and orbital degrees of freedom 
in geometrically frustrated magnet \cite{rf:LiVO2-PRL}. The presence of VBS over the metal-insulator crossover region suggests that, in contrast to the conventional picture, electron transfer (itinerancy) might play a certain role in realizing the VBS, as in the orbitally driven Peierls transition in CuIr$_2$S$_4$ and MgTi$_2$O$_4$ \cite{rf:spin-Peierls}.

Very recently, T. Itou $et$ $al.$ demonstrated the absence of pseudo-gap in the superconducting 
EMe$_3$P[Pd(dmit)$_2$]$_2$ by NMR measurement \cite{rf:Itou}, indicating that the pseudo-gap metallic state is 
unique to the present LiV$X$$_2$ system. 
It may be interesting to further suppress VBS from LiVS$_2$ to $T$ = 0 with substitution of S with Se and to
explore the exotic metal formed near possible VBS critical point.
In the LiVS$_{2-x}$Se$_x$ solid solution, we indeed observed a systematic decrease of
magnetic susceptibility anomaly representing the VBS transition upon Se substitution and
a disappearance around $x$ = 0.3, shown in the right inset of Fig.\ref{fig:phase}.
We found, however, that all the Se substituted samples show
a weakly insulating behavior, perhaps due to the disorder effect inherent to the Se substitution,
and the exotic metal phase including superconductivity could not have been explored.

In conclusion, we have identified a crossover from $S$ = 1 Mott insulator to a paramagnetic metal in a series of triangular lattice vanadates, LiV$X_2$ with $X$ = O, S and Se. LiVS$_2$ is located at the crossover region and a paramagnetic metal to valence bond solid (VBS) insulator transition was observed as a function of temperature. We argue the high temperature metallic phase in LiVS$_2$ is a pseudo-gap metal with possible spin singlet correlation, due to the close proximity to the VBS state, which provides a new playground for the novel interplay of strong electron correlation and the geometrical frustration.
The authors are greatful to D.I. Khomskii, Y. Motome and T. Itou for valuable discussions. This work was partly supported by a Grant-in-Aid for Scientific Research (No.16076204) and ``Nanotechnology Support Project" of the Ministry of Education, Culture, Sports, Science and Technology of Japan.


\begin{thebibliography}{00}
\bibitem{rf:structure} The stacking along the $c$-axis is the 1$T$-type (space group: P$\bar{3}$m1) for LiVS$_2$ and LiVSe$_2$ while the 3$R$-type (space group: R$\bar{3}$m) for LiVO$_2$ in the Ramsdell notation.
\end{thebibliography}
\end{document}